\documentclass[aps,prl,reprint,groupedaddress,showpacs,longbibliography]{revtex4-1}

\usepackage[latin1]{inputenc}
\usepackage[T1]{fontenc}
\usepackage{amssymb,amsmath,bm} 
\usepackage{graphicx}
\usepackage[colorlinks=true,  urlcolor=blue,  linkcolor=blue,  citecolor=blue]{hyperref}
\usepackage{multirow}
\usepackage{siunitx}

\newcommand{\ket}[1] {\left\vert #1 \right\rangle}
\newcommand{\mum}{\rm \si{\micro\meter}}
\DeclareMathOperator{\e}{e}
\newcommand{\braket}[2] {\langle #1 | #2 \rangle}

 \newcommand{\Ust}{U}
 \newcommand{\Hst}{\hat H }
\newcommand{\bs}{\boldsymbol}

\begin{document}

\title{Precision Measurements of Atom-Dimer Interactions in a Uniform Planar Bose Gas}

\author{C. Maury, B. Bakkali-Hassani, G. Chauveau, F. Rabec, S. Nascimbene, J. Dalibard, J. Beugnon}

\affiliation{Laboratoire Kastler Brossel,  Coll\`ege de France, CNRS, ENS-PSL University, Sorbonne Universit\'e, 11 Place Marcelin Berthelot, 75005 Paris, France}

\date{\today}

\begin{abstract}
Cold quantum gases, when acted upon by electromagnetic fields, can give rise to samples where isolated atoms coexist with dimers or trimers, which raises the question of the interactions between these various constituents. Here we perform microwave photoassociation in a degenerate gas of  $^{87}$Rb atoms to create weakly-bound dimers in their electronic ground state. 
From the density-induced shift of the photoassociation line, we measure the atom-dimer scattering length for the two least-bound states of the molecular potential. We also determine the complete energy diagram of one hyperfine manifold of the least-bound state, which we accurately reproduce with a simple model.
\end{abstract}

\maketitle

Cold atomic gases constitute a unique platform to study many-body phenomena. They offer the possibility to relate macroscopic observables, e.g. the equation of state of the fluid, to microscopic quantities such as two- and three-body interactions, as illustrated by the introduction of the ``contact parameter" \cite{Pitaevskii16}. More generally, with a suitable control of these few-body interactions, quantum gases can host simultaneously isolated atoms and molecules, opening intriguing connections with quantum chemistry \cite{Viverit00,iskin2008fermi,Greene17,Bohn17,liu2018building,cheuk2020observation}.

The coexistence of atoms and dimers in a quantum gas raises the question of a universal description of their interactions. In a spin 1/2 Fermi gas close to the unitary limit, the knowledge of the atom-atom scattering length $a$ is sufficient to predict the scattering length $a_{ad}$ characterizing the interaction between an atom and a weakly-bound dimer \cite{skorniakov1957three,Petrov03,Mora04,iskin2008fermi,Levinsen09,Alzetto10,iskin2010dimer,Levinsen11,Alzetto12,Cui14,Zhang14},
 as well as $a_{dd}$, the scattering length for dimer-dimer interaction \cite{petrov2004weakly}.
In a Bose gas in the vicinity of a scattering resonance, the search for a universal relation between $a$ and $a_{ad}$ is more subtle due to the  Efimov effect \cite{naidon_efimov_2017}, i.e. the existence of a large number of three-body bound states when $a$ increases, which requires  the introduction of the so-called ``three-body parameter" \cite{bedaque1999three,Braaten_03_PhysRevA.67.042706,Petrov04,Braaten06,Gao18}.
Outside a resonance, $a$ is comparable to the range of the potential and the existence of a van der Waals universality relating $a_{ad}$ and $a$ for weakly-bound dimers remains an open question \cite{Giannakeas17,Mestrom17,Greene:2022}.

Experimentally, most studies of atom-dimer interactions in quantum gases concentrated so far on inelastic scattering  \cite{Staanum06,Zahzam06,Zenesini14,Lompe10,Nakajima10, Hummon11,Bloom13,Kato17,Yang19,Gregory21,Jurgilas21} and atom-exchange reactions \cite{Knoop10,Rui17,Liu19,Nan19}. Elastic collisions have been studied in a Fermi mixture, in which several partial waves contributed to the scattering process \cite{Jag14} and, more recently, in the context of sympathetic cooling of a molecular gas \cite{Son20}. In this Letter, we concentrate on pure s-wave interactions between atoms and dimers in a rubidium Bose gas. The dimers are prepared either in the least or second-to-least rovibrational bound state using microwave photoassociation. We provide the first precise spectroscopic measurement of the scattering length $a_{ad}$, using a  uniform atomic gas to minimize inhomogeneous broadening of the signal. We complete these results by measuring the full Zeeman diagram of the relevant hyperfine manifold for the least-bound state. 
\begin{figure}[t]
\centering
\includegraphics[width=8.6cm]{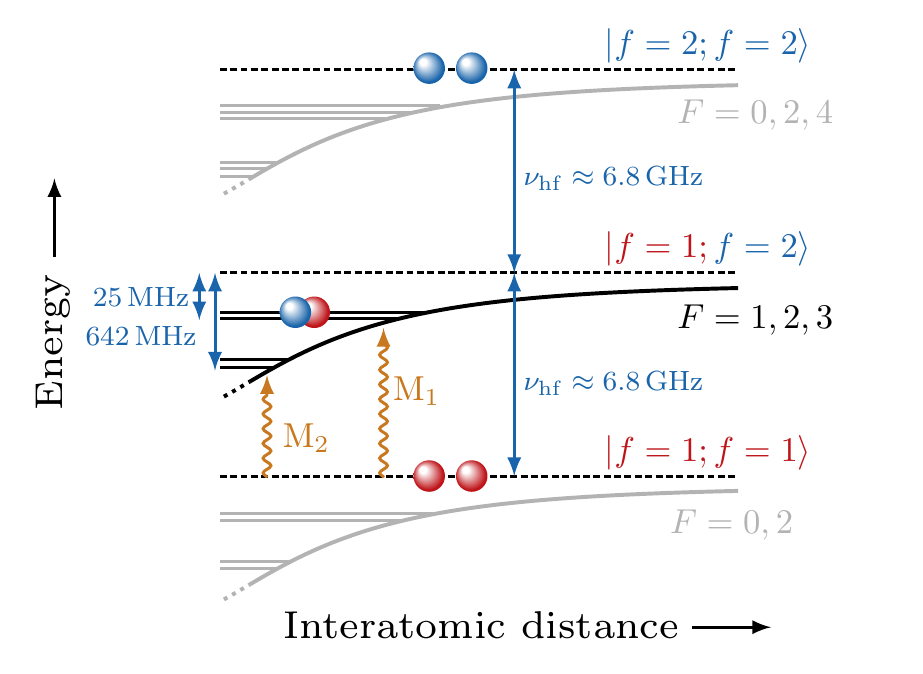}
\caption{Relevant levels for atom pairs in either the  $\ket{f=1}$ or the $\ket{f=2}$ hyperfine sublevels of the electronic ground state of $^{87}$Rb. The dissociation limits of molecular state manifolds, represented as dashed lines, are separated by $h \nu_{\rm hf}\approx h\times 6.8\,$GHz. The molecular potentials are represented by thick continuous lines, together with the values of the total spin angular momentum $F$ of the dimer. Here, we focus on the least-bound rovibrational levels $n=-1$ and $n=-2$ of the $\ket{f=1;f=2}$ branch, with zero orbital angular momentum. These levels are located  at  $\sim 25$ and 642\,MHz below the dissociation energy, respectively.  The dimers are produced by microwave photoassociation of two atoms either both in $\ket{f=1}$ or both in $\ket{f=2}$. }\label{fig:levels}
\end{figure}

We use uniform, horizontal flat Bose gases of $^{87}$Rb atoms at low temperature ($T\lesssim20\,$nK) prepared in the strongly-degenerate regime \cite{Ville17}. 
The confinement along the vertical direction $z$, provided by a harmonic potential of frequency $\omega_z/2\pi \approx 3.7$\,kHz, is strong enough to ensure that $k_{\rm B}T,\mu_a \ll \hbar\omega_z$, where $\mu_a$ is the chemical potential of the gas. This ensures the two-dimensional character of the fluid at the thermodynamic level. 
However, the thickness of the cloud along $z$, $\ell_z=\sqrt{\hbar/m\omega_z}\approx\,180\,$nm, remains much larger than the scattering length $a\approx\,5\,$nm, so that collisions keep a three-dimensional character \cite{Petrov:2000a}.
Unless otherwise stated, the in-plane confinement is a disk-shaped box potential of radius 20\,$\mum$.

\begin{figure}
\centering
\includegraphics[width=8.6cm]{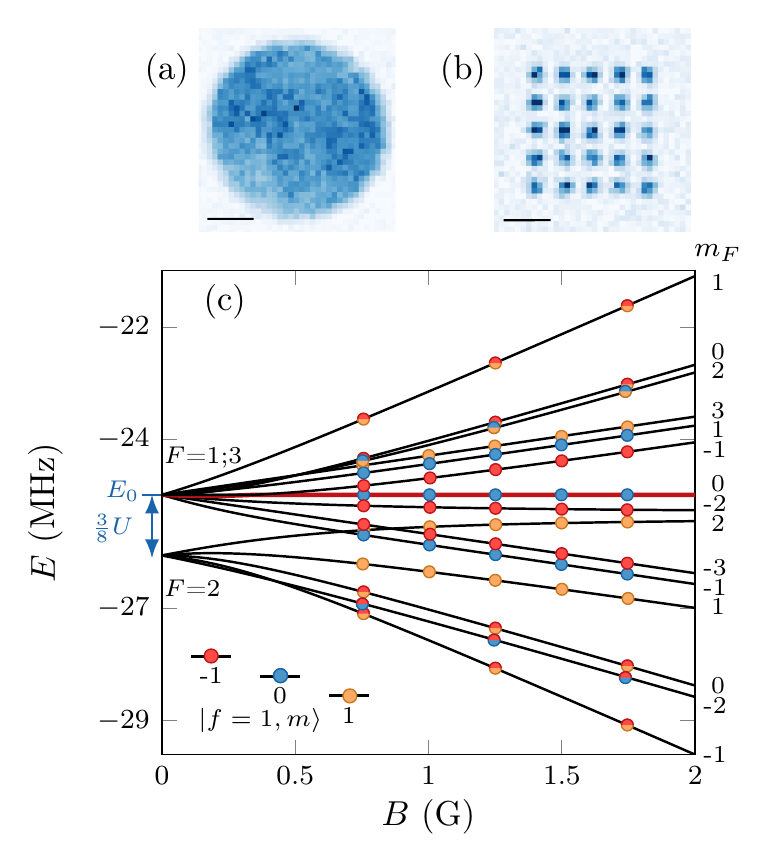}
\caption{(a-b) Clouds used for microwave spectroscopy with 
either (a) a single component state $\ket{f=1,m=0,\pm1}$ or (b) a binary mixture of these hyperfine states (bar length: 10$\,\mum$). (c) Energy diagram of the  manifold ${\cal E}^{(n=-1)}_{12}$. The colors of the experimental points encode the composition of the initial atomic state. The set of solid lines is the result of the simple model described in the text, with two adjustable parameters $U$ and $E_0$.  The $|\Psi_0^{(n=-1)}\rangle$ used for characterizing atom-dimer scattering is highlighted with a red solid line. The numbers on the right give for each state the $z$-component $m_F$ of its total angular momentum.}
\label{fig:diag}
\end{figure}

For $^{87}$Rb, the electronic ground level is split by hyperfine interaction into two sublevels $f=1$ and  $f=2$, separated by an energy difference $h \nu_{\rm hf}\approx h \times 6.8\,$GHz. In most experiments described below, the atoms are prepared in the state $\ket{f=1, m=0}$ with the quantization axis along $z$ defined by a magnetic field $B$ in the range 0.7 to 2\,G \footnote{For some molecular transitions of the Zeeman diagram in Fig.\,\ref{fig:diag} the field is oriented in the horizontal plane to maximize the coupling to the molecular state.}. The relevant energy levels for a pair of atoms are sketched in Fig.\,\ref{fig:levels}, together with the molecular potentials leading to the formation of dimers. 

Starting with a gas in the hyperfine level $f=1$, we shine a microwave field of typical amplitude $B_{\rm mw}\approx 30$\,mG  to photoassociate atom pairs into weakly-bound dimers of the $\ket{f=1 ; f=2}$ manifold. We first target the least-bound  ($n=-1$) rovibrational level with zero orbital angular momentum and a binding energy  $\sim - h\times 25$\,MHz \cite{Freeland01}. This level has a hyperfine structure associated to three possible values of the total spin angular momentum $F=1,2,3$. More specifically, within this multiplicity of dimension 15, we focus on a specific state, labeled $|\Psi_0^{(n=-1)}\rangle$. Its spin component is $(\sqrt{3} |F=3,m_F=0\rangle-\sqrt{2}|F=1,m_F=0\rangle)/\sqrt{5}$ in the limit of low magnetic field, it is a pure electronic spin triplet and has zero magnetic moment. We denote  M$_1$ the transition from  $\ket{f=1,m=0 ; f=1,m=0}$ to $|\Psi_0^{(n=-1)}\rangle$. It is well suited for precision measurements of interaction energies because it is insensitive to magnetic field fluctuations at first order. 

\begin{figure}[t]
\centering
\includegraphics[width=8.6cm]{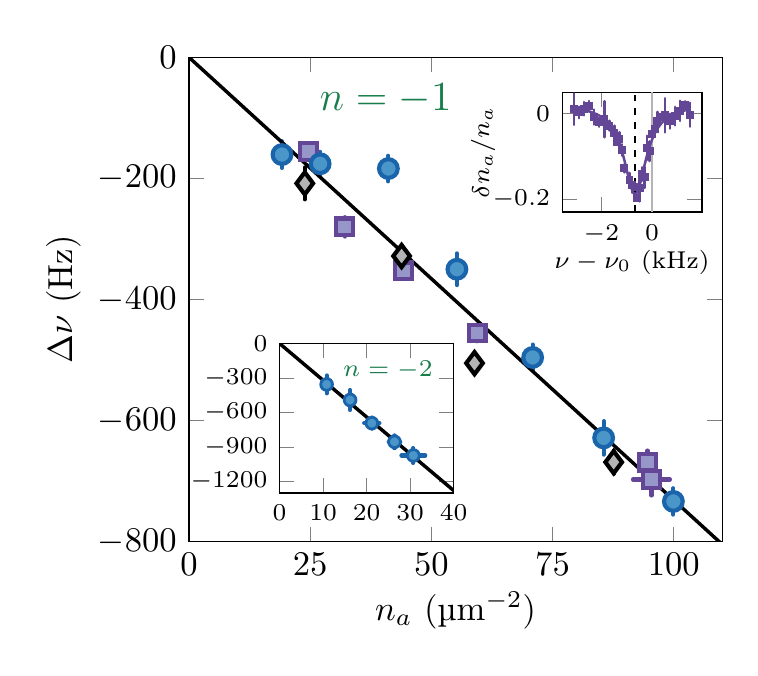}
\caption{Frequency shift of the M$_1$ line as a function of the surface density of the gas.  The three different symbols correspond to a measured depletion signal of 20\% (square), 14\% (diamond), 8\% (circle). 
All data are adjusted by a common linear fit. The left-hand inset shows the result of a similar measurement for the second-to-least bound level (M$_2$ line in Fig.\,\ref{fig:levels}). The right-hand  inset shows a typical microwave photoassociation signal for $n_a=95$\,$\mum^{-2}$. The variation of the width of the photoassociation signal with density is reported in \cite{REFSM}.} \label{fig:shift}
\end{figure}

We detect the formation of dimers by losses in the gas as a function of the microwave frequency $\nu$. A typical  signal on the M$_1$ transition is reported in the upper right inset of Fig.\,\ref{fig:shift}. The reference frequency $\nu_0$ corresponds to the resonance frequency measured in the zero-density limit \footnote{$\nu_0$ differs from $\nu_{\rm hf}$ because of the second-order Zeeman effect for the atom and dimer states.}. For this reported spectrum, the peak frequency $\nu_m$ is displaced with respect to $\nu_0$ by $\approx 670\,$Hz and the measured full width at half maximum is $ \approx 1$\,kHz.   We attribute it to the finite lifetime of the dimers, which can decay through two channels. First, dipolar relaxation within the dimer can produce a pair of atoms by releasing the energy $h \nu_{\rm hf}$ \footnote{The lifetime of the molecules is much larger than the naive estimate obtained from the known two-body loss rate ( $ \sim 10^{-14}$\,cm$^3.$s$^{-1}$) for a pair of atoms in $f=1$ and $f=2$ and confined in a volume given by the size of the dimers, i.e. $a^3\sim 100$\,nm$^3$, see also \cite{verhaar2017stability}}. Second, two-body inelastic collisions between atoms and molecules can play a significant role, as observed for other alkali molecules \cite{Staanum06,Zahzam06,Zenesini14,Mordovin15}.  

To study the interaction between the produced dimers and the atom gas, we measure the variation of the peak frequency $\Delta \nu\equiv \nu_m-\nu_{0}$ with the surface atomic density $n_a$. For each density, we adjust the duration of the excitation time to keep the depletion at resonance $\delta n_a/n_a$ at a given value. We show the results obtained  for three different values of $\delta n_a/n_a$ in Fig.\,\ref{fig:shift}. We observe a shift of the resonant frequency which goes up to 800\,Hz at our maximum density $n_a\sim$\,100 atoms/$\mum^2$. All data collapse on a single curve, which confirms that we operate in the weak excitation regime. We fit a linear function to the data and obtain  $\Delta \nu/n_a=-7.3(3)$\,Hz/$\mum^2$ \footnote{The data were plotted after subtracting the intercept from the linear fit so that the reference frequency corresponds to the zero-density limit.}.  

In order to interpret this shift within a mean-field approach, we introduce the interaction parameter $g_{ad}=2\pi a_{ad} \hbar^2/m_r$, where $m_r=2m_a/3$ is the reduced mass of the atom-dimer system. We assume that all interactions occur in the  s-wave regime, because of the very low relative momenta between the unbound atoms and the dimer. The photoassociation process must bring to the sample (i) the energy in the zero density limit $h\nu_{0}$, (ii) the interaction energy between the dimer and the atom bath, (iii) the energy $-2\mu_a$, since two atoms are removed from the bath. Denoting $\rho_a ({z})$  the 3D density profile of the atom bath and $f_d ({z})$ the distribution function of the dimer (with the normalization $\int \mathrm{d} z\, \rho_a ({z})=n_a$ and $\int \mathrm{d} z\, f_d ({z})=1$), we find using the mean-field value of $\mu_a$ in the low temperature limit \cite{REFSM}
\begin{equation}
h\Delta \nu =g_{ad} \int \mathrm{d} z\, \rho_a ({z})f_d ({z})-2\mu_a=\left(\frac{\sqrt 3\, a_{ad}}{2a_1}-2 \right)\mu_a, 
\label{eq:shiftdimer}
\end{equation}
where $a_f$ denotes the s-wave scattering length of an atom bath in state $\ket{f,m=0}$ ($f=1$ or 2). In all cases the dimer density is low enough so that dimer-dimer interactions can be safely neglected.

The atomic surface density $n_a$, or equivalently the chemical potential $\mu_a$, are inferred via Ramsey spectroscopy. We measure  the density-dependent component $\Delta \nu'$ of the microwave frequency that allows a full transfer of the gas from $|f=1,m=0\rangle$ to $|f=2,m=0\rangle$ and find $\Delta \nu'/n_a= -0.52(2) $\,Hz/$\mum^2$.  Combining Eq.\eqref{eq:shiftdimer} with 
\begin{equation}
h\Delta \nu'=\frac{1}{2}\frac{a_2-a_1}{a_1}\mu_a,
\label{eq:shiftatom}
\end{equation}
we obtain the atom-dimer scattering length:
\begin{equation}
a_{ad}= \frac{4}{\sqrt{3}}a_1+\frac{1}{\sqrt{3}} \frac{\Delta \nu}{\Delta \nu'} (a_2-a_1),  \label{eq:aad}
\end{equation}
an expression which is immune to systematic errors in the calibration of the density $n_a$. 
Using the known values of $a_1=100.9\,a_0$ and $a_2-a_1=-6\,a_0$ with $a_0$ the Bohr radius \cite{Altin11}, we obtain 
\begin{equation}
a_{ad}^{(n=-1)}[\mathrm{bath\: in\,} f=1]=184(2)\,a_0,
\end{equation} 
where the quoted error  takes into account only the uncertainties on $\Delta \nu$ and $\Delta \nu'$ \footnote{The theoretical predictions of \cite{Altin11} for $a_f$ are quoted with no error bars. Previous experiments in our group \cite{Zou20a,Zou21a} confirmed the prediction for $a_1-a_2$ with a $\sim 10\%$ accuracy. A shift of $a_1-a_2$ by this amount would lead to a shift of $a_{ad}$ by a few times the reported uncertainties.}. This result is notably different from the result $a_{ad}^{\rm impulse}=8a_1/3=269\,a_0$ of the impulse approximation \cite{Chew52}, which consists in summing independently the scattering amplitudes of an atom of the bath with each atom of the dimer.

We have performed two additional measurements of the atom-dimer scattering length, by changing either the final or the initial state of the photoassociation process: \\
\noindent $-$ Firstly, still starting from a gas of atoms in $\ket{f=1,m=0}$, we produced dimers in the state  $|\Psi_0^{(n=-2)}\rangle$ of the second-to-least bound rovibrational level (M$_2$ transition in figure \ref{fig:levels}). The target state still has zero orbital angular momentum and it is the magnetic-field-insensitive  state equivalent to that studied above for the $n=-1$ multiplicity. We found the value $h\times - 642.219(1)$\,MHz for its binding energy at zero magnetic field, and the value
 \begin{equation}
a_{ad}^{(n=-2)}[\mathrm{bath\: in\,} f=1]=21(7)\,a_0
\end{equation} 
 for the atom-dimer scattering length, see inset of Fig.\,\ref{fig:shift}. 
The large difference between $a_{ad}^{(n=-1)}$ and $a_{ad}^{(n=-2)}$ shows the key role of the dimer radial wavefunction in the scattering process. 
We note that a related work was performed with two-photon photoassociation of $^{87}$Rb atoms for a state in the $n=-2$, $\ket{f=1;f=1}$ manifold  and trapped in a harmonic potential \cite{Wynar00}. An atom-dimer scattering length of $-180(150)\,a_0$ was reported, where the large uncertainty could be attributed to the difficulty of accurately modeling  the experimental signal in an inhomogeneous cloud.\\
\noindent $-$ Secondly, coming back to the dimer state $|\Psi_0^{(n=-1)}\rangle$, we  measured its interaction with a bath of atoms  initially all in state $\ket{f=2,m=0}$. 
The fitted slope is now $8.1(9)$\,Hz/$\mum^2$, leading to 
\begin{equation}
a_{ad}^{(n=-1)}[\mathrm{bath\: in\,} f=2]=165(7)\,a_0,
\end{equation} 
a value close to the result of an atom bath in state $\ket{f=1,m=0}$. The similarity between these two results, combined with the small difference between the two bath scattering lengths $a_1$ and $a_2$, is compatible with the existence of a ``van der Waals universality", which may allow one to link $a$ and $a_{ad}$ for the least-bound dimers.

We now turn to the detailed analysis of the least-bound manifold ${\cal E}^{(n=-1)}_{12}$ emerging from a pair of atoms in the energy level $\ket{f_1=1;f_2=2}$. We show in Fig.\,\ref{fig:diag}c the measurement of the energy of all 15 states as a function of the applied external magnetic field $B$. By preparing the gas in $\ket{f=1,m=0}$ as described previously, we can photoassociate only three states (blue circles) corresponding to the absorption of a photon with a $\pi$ or $\sigma_\pm$ polarization. The rest of the diagram is obtained by preparing the gas in other pure hyperfine states $\ket{f=1,m=\pm 1}$ or in binary mixtures of $\ket{f=1,m=0,\pm 1}$. Experiments with mixtures require a modification of the experimental protocol: as the two species have different magnetic moments, residual magnetic field gradients lead to spatial separation of the two components and  prevent the dimer formation. We circumvent this problem by loading atoms in an array of microtraps whose size (about 5\,$\mum$) is small enough to prevent phase separation (Fig.\,\ref{fig:diag}ab).

The energy diagram of Fig.\,\ref{fig:diag} was previously computed using coupled channels calculations \cite{Tscherbul10} and quantum defect theory \cite{Hanna10}, and it has already been partially measured \cite{Mordovin15}. Here, we present a simple model that accurately reproduces the experimental data. At zero magnetic field, collisions between two $^{87}$Rb atoms involve two  channels, associated to singlet and triplet potentials with slightly different scattering lengths. We model this difference by the Hamiltonian
\begin{equation}
\hat H_{\rm st}=U\, \hat{\bm{s}}_1 \cdot \hat{\bm{s}}_2=\frac{U}{2}\,\left (\hat{\bm{S}}^2-3/2\right)
\end{equation}
acting as a perturbation term in the subspace $\ket{f=1;f=2}$ of the least-bound level $n=-1$.
Here $\hat{\bm{s}}_i$ designates the electron spin operator of atom $i$, $\hat{\bm{S}}=\hat{\bm{s}}_1+\hat{\bm{s}}_2$, and $U$ is an adjustable parameter. Noting that states with well-defined values of $F$ are eigenstates of $\hat H_{\rm st}$, we find for $B=0$ the first-order energy shifts $\Delta E_{F=1}=\Delta E_{F=3}=-2\Delta E_{F=2}=U/4$  \cite{REFSM}. 

The effect of the magnetic field is described by $\hat H_Z \approx 2 \mu_B B \hat S_z$, which mixes all $F$ states. 
The diagonalization of the 36x36 matrix of the spin Hamiltonian $\hat H_{\rm st} + \hat H_Z$  leads for the manifold $\mathcal{E}^{(n=-1)}_{12}$ to the 15 continuous lines shown in the energy diagram of Fig.\,\ref{fig:diag}. These lines are obtained by adjusting two parameters: the coupling $U$ and the energy $E_0\equiv E_{F=1}(B=0)$.  We obtain $U=h\times2.875(5)$\, MHz and  $E_0=h\times -24.985(1)$\,MHz. This model provides an excellent agreement with all 15 lines of the measured diagram. The distance between the best fitted energy diagram and our measurements is $3.7$~kHz \footnote{the distance between the data and the fit is defined as $[\sum_i \left(\nu_i^{\rm (meas)}-\nu_i^{\rm (fit)}\right)^2/N_{\rm points}]^{1/2}$, where  $\nu_i^{\rm (meas)}$ and $\nu_i^{\rm (fit)}$ are the measured and fitted frequency, respectively, and $N_{\rm points}$ is the number of measured points}, which is slightly larger than the uncertainty on the measured molecular line positions ($\lesssim 1\,$kHz).  The remaining deviations remind us that this simple model is not expected to be exact. A similar study could be performed to measure the Zeeman diagram of the $\ket{f=1;f=1}$ and $\ket{f=2;f=2}$ multiplicities. In our case, the limited lifetime of the atomic sample prepared in $\ket{f=2}$, comparable to required excitation times, makes these experiments challenging.

In conclusion, we have presented a precise measurement of the scattering length $a_{ad}$ characterizing the interaction between atoms and weakly-bound dimers in a degenerate Bose gas. This result provides a first step in the search for a possible van der Waals universality for this problem. 
Our method can be straightforwardly adapted to other alkali-metal bosonic species. Some of them provide easily accessible Feshbach resonances, making it possible to study the emergence of Efimov physics on $a_{ad}$. 
In addition, our precise determination of the whole energy diagram of a weakly-bound dimer manifold paves the way to the implementation of  microwave Feshbach resonances  \cite{Papoular10,Ding17}, using for instance strong microwave fields directly generated on atom chips \cite{Bohi09}.

\begin{acknowledgments}
This work is supported by ERC  (grant agreement No 863880) and the ANR-18-CE30-0010 grant. We thank O. Dulieu, F. Chevy, C. Greene, J. D'Incao, S. Kokkelmans and D. Papoular for fruitful discussions.
\end{acknowledgments}

\bibliography{Molecules_bib}

\appendix

\clearpage

\renewcommand{\thefigure}{S\arabic{figure}} 
\setcounter{figure}{0}

\section{SUPPLEMENTAL MATERIAL}

\section{Photoassociation spectroscopy }
We detail in this section the measurements of the M$_2$ line reported in the main text and the additional study of photoassociation from $\ket{f=2; f=2}$ to  $|\Psi_0^{(n=-1)}\rangle$. We note that precision measurement of $a_{ad}$ for other states of the same rovibrational level is not accessible in our setup because most other lines present a first-order Zeeman effect and our residual magnetic field noise ($\sim$ mG) thus blurs the signal. The other magnetic field insensitive transitions are  too weak to allow an accurate measurement. 

\subsection{M$_2$ transition}
We describe here the parameters used to obtain the data reported in the inset of Fig.\,2 in the main text for the  M$_2$ transition from $|f=1,m=0;f=1, m=0\rangle$ to $|\Psi_0^{(n=-2)}\rangle$. The strength of this transition is much weaker than the one of the M$_1$ transition due to a lower overlap of the radial wavefunctions.  We thus use 10\,s pulses compared to duration $< 260\,$ms for the M$_1$ transition.  Because of this longer duration, we had to decrease the microwave power to $0.76\,P_\mathrm{max}$ to limit thermal effects in the microwave amplifier. Finally, we restrict ourselves to densities below $30\,\mum^{-2}$ for which the observed shift is linear with density. For higher densities, the change of density during the probing duration makes the interpretation of the data more involved. In the inset of Fig.\,2 of the main text, the horizontal error bars indicate the density range scanned during excitation.

\subsection{Width of M$_1$ and M$_2$  transitions}
\begin{figure}[ht!]
\centering
\includegraphics[width=7.5cm]{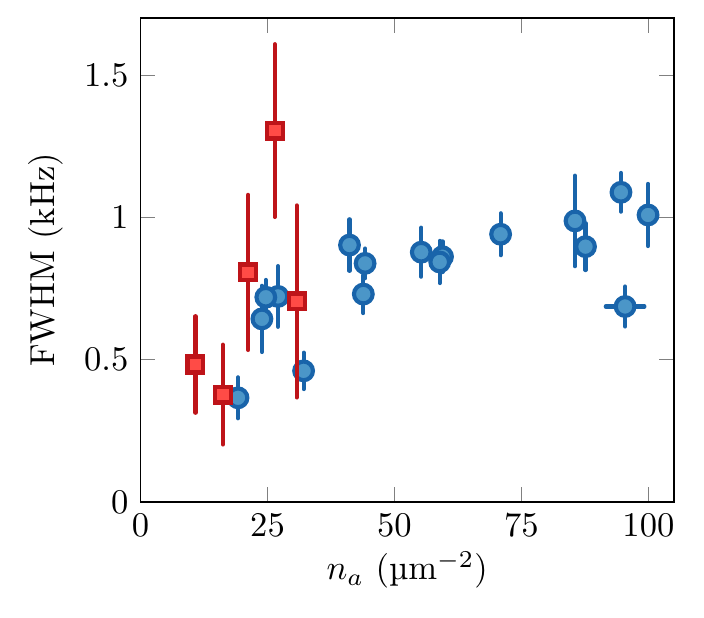}
\caption{FWHM of the M$_1$ (blue circles) and M$_2$ (red squares) lines as a function of the surface density of the atomic cloud for the same data as the one reported in Fig.\,2 of the main text.}\label{fig:width_M1}
\end{figure}

We show in Fig.\,\ref{fig:width_M1} the measured full width at half maximum (FWHM) of the loss spectroscopy signals used for determining the position of the M$_1$ and M$_2$ lines as a function of atomic density. For both lines, the width is always $\lesssim1$\,kHz and we observe an increase of the linewidth with density. There is no significant difference between the two lines.

\subsection{Transition from $\ket{f=2; f=2}$}
\begin{figure}[ht!]
\centering
\includegraphics[width=7.5cm]{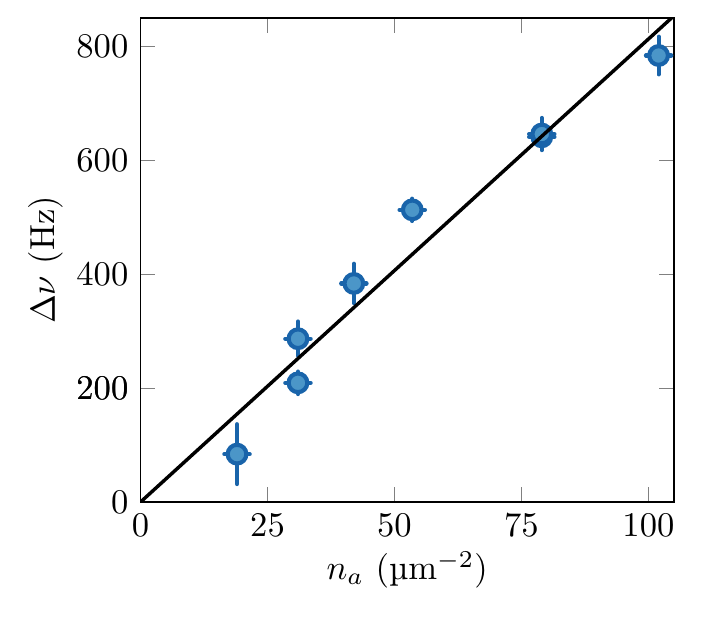}
\caption{Frequency shift of the $|f=2,m=0;f=2, m=0\rangle\leftrightarrow|\Psi_0^{(n=-1)}\rangle$ transition as a function of the surface density of the atomic cloud. The microwave power is set to its maximal value $P_\mathrm{max}$. For all points, the duration of excitation is $< 10$~ms and is adjusted to keep the maximal depletion constant around 8$\,\%$. The  solid black line is a linear fit to the data.}\label{fig:shift_f2}
\end{figure}

The data reported in Fig.\,2 of the main text correspond to photoassociation spectroscopy from a $|f=1,m=0;f=1, m=0\rangle$ unbound state to the $|\Psi_0^{(n=-1)}\rangle$ and $|\Psi_0^{(n=-2)}\rangle$ molecular states. We show in Fig.\,\ref{fig:shift_f2} the results of photoassociation from a $|f=2,m=0;f=2, m=0\rangle$ unbound state to the $|\Psi_0^{(n=-1)}\rangle$ state. These measurements are limited by the short lifetime of the atomic sample in $|f=2,m=0\rangle$ of $\lesssim 100\,$ms for the explored densities. However, the strength of this transition is stronger than the one starting from  $|f=1,m=0;f=1, m=0\rangle$, which makes the signal large enough for probing times as short as $\sim\,10\,$ms. We also observe a linear shift of the line center with the density. A linear fit to the data gives a slope of $8.1(9)$\,Hz/$\mum^2$. 

\section{Modeling of the photoassociation signal}
The dynamics of photoassociation in cold Bose gases has been for instance discussed in Ref.\,\cite{Naidon08}. In the case of dilute cold gases, where the molecules are lost from the system, the time evolution of the atomic density in a two-body model is expected to follow the rate equation
\begin{equation}
\frac{\mathrm{d}\rho_a}{\mathrm{d}t}=-\beta \rho_a^2,
\end{equation}
where $\beta$ is the time-independent loss rate coefficient. This result also applies to the many-body situation of dense gases when the molecule loss rate is large enough \cite{Naidon08}. This contrasts with the case with low losses and strong excitation, leading to coherent coupling between the atomic and molecular states, as observed for example in Ref.\,\cite{Yan13} with Strontium atoms.

In all experiments reported in this Letter,  we chose short enough excitation duration to remain in a regime in which the bath depletion is small and our excitation strength is low enough so that we do not observe coherent oscillations between the atomic and molecular states. We thus describe the photoassociation process via a Fermi golden rule approach. In this regime the variation of the 3D atomic density during a time $\delta t$ is given by
\begin{equation}
\delta \rho_a \propto - \beta \rho_a^2 \delta t.
\end{equation}
The loss rate coefficient is proportional to $|\braket{i}{\hat V |f}|^2$, where $\hat V$ is the operator describing the coupling induced by the microwave field and the indices $i,f$ describe the initial atomic and final molecular states, respectively. In our experiments we start with an atomic cloud whose 3D density is uniform in the $xy$ plane and given by the probability distribution of the ground state of an harmonic oscillator of size $\ell_z=\sqrt{\hbar/m\omega_z}$ along the vertical direction, which thus writes
\begin{equation}
\rho_a(z)=\frac{n_a}{\ell_z \sqrt{\pi}} \e^{-z^2/\ell_z^2}.
\end{equation}
The initial two-atom state can be decomposed in its center-of-mass and relative motion states
\begin{equation}
\ket{i}=\ket{\boldsymbol{K},0}\otimes \ket{\phi_{\boldsymbol{k}}},
\end{equation}
where $\boldsymbol{K}$ is the center-of-mass in-plane momentum, and the index 0 refers to the harmonic oscillator ground state along the $z$ direction. The relative motion of the two atoms in our quasi-2D system is described by the scattering state $\ket{\phi_{\boldsymbol{k}}}$ as defined in Ref.\,\cite{Petrov01}. We note that the frequency of the harmonic trap is larger than the typical linewidth of the photoassociation lines, which justifies to use the vibrational state basis to describe the initial and final states of the photoassociation process. The final molecular state can also be decomposed as a center-of-mass and relative motion state
\begin{equation}
\ket{f}=\ket{\boldsymbol{K},n_d}\otimes \ket{\phi_{d}}.
\end{equation}
The center-of-mass in-plane momentum $\boldsymbol{K}$ is unchanged by the microwave photon. We have introduced $n_d$, the vibrational number associated to the molecular center-of-mass state of the vertical harmonic oscillator.
We note that the relative motion at the scale of the extension of the molecular state is not influenced by the confinement along $z$ because the molecule extension is much smaller than $\ell_z$.
We now assume that the oscillation frequency $\omega_z$ is the same for the atom and the dimer center-of-mass, which corresponds to a trapping force twice larger for the dimer whose mass is twice the atomic mass. As the microwave field does not influence the center-of-mass motion, this leads to $n_d=0$. This assumption is supported experimentally by the absence of vibrational sidebands in the measured photoassociation spectrum.
One can then determine $f_d$, the  distribution function of the center-of-mass of a single dimer. It is uniform in-plane and depends on its vertical coordinate $Z$ as
\begin{equation}
f_d(Z)=\frac{\sqrt{2}}{ \ell_z \sqrt{\pi}}\e^{-2 Z^2/\ell_z^2}.
\end{equation}
Because the formation of a dimer requires two atoms close to each other, the density distribution of molecules is indeed narrower than the atomic one along the $z$ direction. From this expression of $f_d$, we thus get 
\begin{equation}
 \int \mathrm{d}z\, \rho_a ({z})f_d ({z})=\sqrt{\frac{2}{3\pi}}\frac{n_a}{\ell_z},
\end{equation}
which is used in Eq.\,(1) of the main text with the definition of the chemical potential for a quasi-2D gas, $\mu_a= \frac{\hbar^2}{m}\sqrt{8\pi}\frac{a}{\ell_z}n_a$,  to obtain the reported expression of $\Delta \nu $.

\section{Zeeman diagram: theory}

\subsection{Hyperfine structure at $B=0$}
We consider a weakly-bound rovibrational state of the Rb$_2$ dimer with $\ell=0$ and focus on its hyperfine spectrum. The state of each atom of the dimer can be decomposed in the $\{f,m_f\}$ basis of dimension 8 ($f=1,2$). This leads to 64 possible two-atom spin states, but only 36 are symmetric under particle exchange as required for symmetric orbital wavefunctions. These states can be split into three multiplicities ${\cal E}_{1,1} \equiv(f=1,f=1)$, ${\cal E}_{1,2} \equiv(f=1,f=2)$ and ${\cal E}_{2,2} \equiv(f=2,f=2)$, separated in energy by the atomic hyperfine splitting of the electronic ground state of about $\nu_{\rm hf}\sim 6.8$\,GHz. We focus in the main article only on multiplicity  ${\cal E}_{1,2} $, but we give here some information about multiplicities ${\cal E}_{1,1} $ and ${\cal E}_{2,2} $.

The possible values of the total angular moment $F$ of the dimer state are thus $F=0,1,2$ for multiplicity ${\cal E}_{1,1} $,  $F=1,2,3$ for multiplicity ${\cal E}_{1,2} $  and  $F=0,1,2,3,4$ for multiplicity ${\cal E}_{1,2} $. However, only even values of $F$ corresponds to states that are symmetric under exchange for multiplicities ${\cal E}_{1,1} $ and ${\cal E}_{2,2} $.

For $^{87}$Rb, the singlet and triplet potentials lead to similar, but not exactly equal, scattering lengths. To take into account this difference for the least bound states, we add a phenomenological term to the Hamiltonian
\begin{equation}
\Hst= \Ust\;\hat{\bs s}_1\cdot \hat{\bs s}_2 =\frac{\Ust}{2}\left( \hat {\bs S}^2-\frac{3}{2}\right)
\end{equation}
where $\hat {\bs S}=\hat{\bs s}_1+ \hat{\bs s}_2$ stands for the total electronic spin.

\begin{figure}[ht!]
\begin{center}
\includegraphics[width=8.6cm]{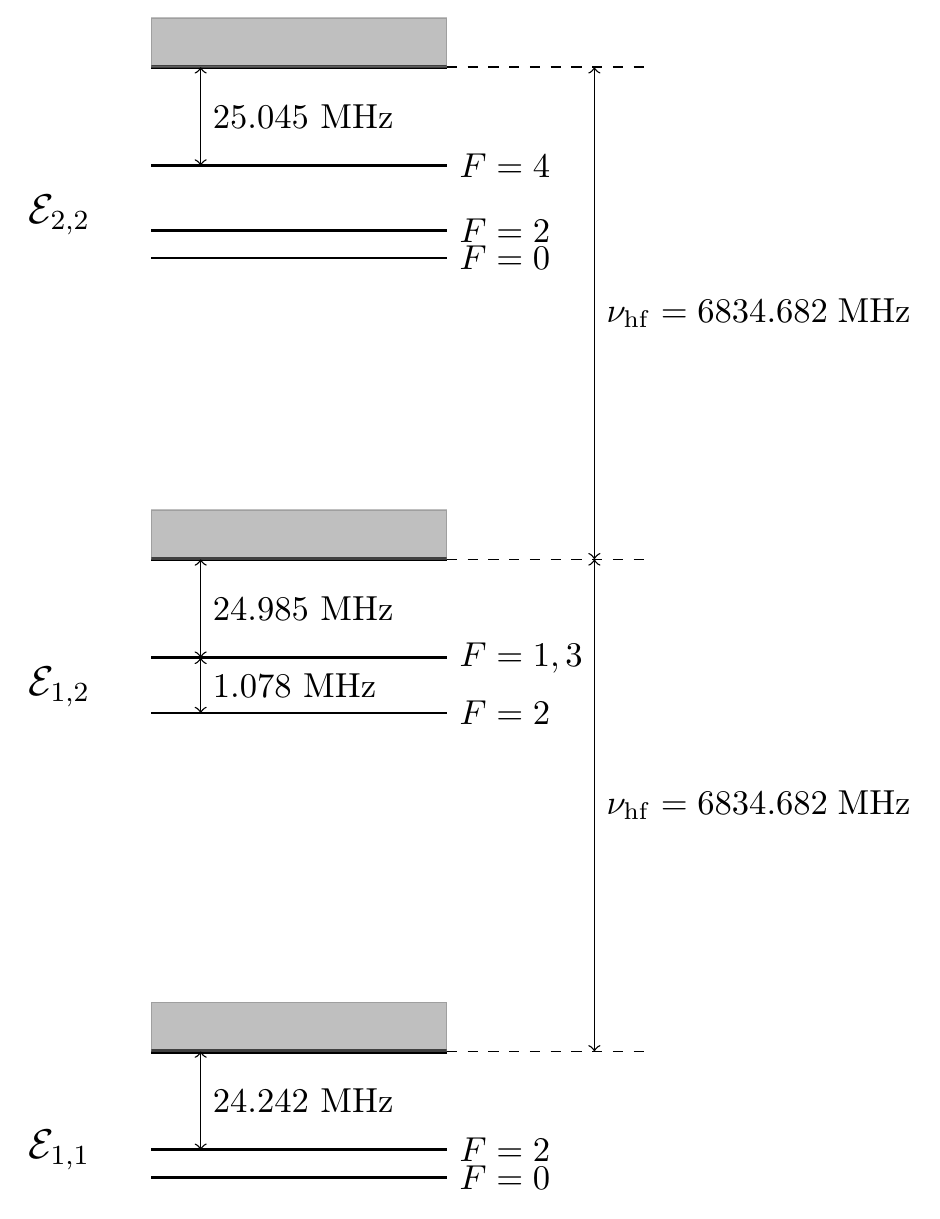}
\end{center}
\caption{Spectrum for a zero magnetic field. The values quoted for the multiplicities $\mathcal{E}_{1,1}$ and $\mathcal{E}_{2,2}$ are extracted from Ref.\,\cite{Freeland01}. The values reported in this work for multiplicity $\mathcal{E}_{1,2}$ are in excellent agreement with those quoted in \cite{Mordovin15}. }
\label{fig: spectrum}
\end{figure}

Since $\Ust\ll h\nu_{\rm hf}$, we use degenerate perturbation theory and diagonalize the restriction of $\Hst$ inside each multiplicity ${\cal E}_{1,1} ,{\cal E}_{1,2} ,{\cal E}_{2,2} $. The total angular momentum $F$ remains a good quantum number because $\hat {\bs S}^2$ commutes with the total electronic $+$ nuclear angular momentum $\hat {\bs F}$.  We obtain
\begin{itemize}
 \item
Multiplicity ${\cal E}_{1,1}$: 
\begin{equation}
\Delta E_{F=0}=-2\Delta E_{F=2}=-U/8
\end{equation}

 \item
 Multiplicity ${\cal E}_{1,2}$:
 \begin{equation}
\Delta E_{F=1}=\Delta E_{F=3}=-2\Delta E_{F=2}=U/4
\end{equation}
Note that the subspaces $F=1$ and $F=3$ are pure electronic-spin-triplet, corresponding to $S=1$. 

\item
Multiplicity ${\cal E}_{2,2}$:
\begin{equation}
\Delta E_{F=0}=2\Delta E_{F=2}=-3\Delta E_{F=4}/2=-3U/8
\end{equation}
The subspace $F=4$ is pure electronic-spin-triplet.

\end{itemize}
The situation is summarized in Fig.\,\ref{fig: spectrum} where we indicate the spectroscopic values known so far. Note that the value of $\Ust$ may depend on the multiplicity. 

\subsection{Zeeman effect}
The coupling to an external magnetic field aligned with the $z$ axis is described by the Zeeman Hamiltonian
\begin{equation}
\hat H_{\rm Z}=\mu_B B (g_e \hat S_z+ g_i \hat I_z)
\end{equation}
with the electronic and nuclear Land\'e factors $g_e=2.002319$ and $g_i=-0.000995$. We note that $\hat H_{\rm Z}$ is a rank 1 operator so it cannot couple two subspaces differing by $\Delta F\geq 2$. Therefore, $\hat H_{\rm Z}$ just leads to a linear variation of the energy levels with the magnetic field inside the multiplicities $\mathcal{E}_{1,1}$ and $\mathcal{E}_{2,2}$, with no avoided crossing. In the multiplicity $\mathcal{E}_{1,2}$, one has to diagonalize it numerically. The result of this diagonalization is shown in Fig.\,3 of the main text.

\subsection{The $|\Psi_0^{(n)}\rangle$ state}
The $|\Psi_0^{(n)}\rangle$ state has, within perturbation theory, zero magnetic moment. It can be written as the symmetric $ |f_1=1,m_1=0;f_2=2,m_2=0\rangle $ state, or
\begin{equation}
|\Psi_0\rangle=\frac{1}{\sqrt{2}}\left(\ket{++;-\frac{1}{2}-\frac{1}{2}}-\ket{--;+\frac{1}{2}+\frac{1}{2}}\right)
\end{equation}
 in the basis $|s_{1z},s_{2z};i_{1z},i_{2z} \rangle$ where $i_{jz}=\pm1/2,\pm3/2$ is the projection of the nuclear spin of atom $j$.

Note that the state $|\Psi_0^{(n)}\rangle$ is a pure electronic spin-triplet and that it does not have a well-defined value of $F$: it is a linear combination of $|F=1,m_F=0\rangle$ and $|F=3,m_F=0\rangle$, namely
\begin{equation}
\sqrt{\frac{3}{5}} |F=3,m_F=0\rangle-\sqrt{\frac{2}{5}}|F=1,m_F=0\rangle.\label{eq:flatstateB}
\end{equation}

Finally, one can check that this state is not coupled by $\Hst$ or $\hat H_{\rm Z}$ to the two other states with $m_F=0$ within the multiplicity $\mathcal{E}_{1,2}$. Therefore its magnetic moment is null at this level of approximation.

\begin{table*}[ht!]
\begin{center}
\begin{ruledtabular}
\begin{tabular}{lccccccccc}

  $m_F$ & $N$ & Initial state & Cloud & P & $B_P$ [mG]  & $t$ [s] & $\delta n_a/n_a$ [$\%$] & FWHM[kHz]&$\Delta \nu$(1.25G) [kHz]\\
  \hline                    
  
  3 & \rule{0pt}{4ex}12 & $|+1; +1\rangle$ & Bulk & $\sigma^+$ & 12 & 5 & 33 & 3.2&-2\\
  
  \hline 
  
  \multirow{2}{*}{2} & \rule{0pt}{4ex}7 & $|+1;+1\rangle$ & Bulk & $\pi$ & 31  & 1 & 20 & 3.6 &-2\\
  & \rule{0pt}{4ex}13 & $|+1; 0\rangle$ & Patches & $\sigma^+$ & 12   & 10 & 17 & 2.9&5\\
  
  \hline 
  
  \multirow{3}{*}{1} & \rule{0pt}{4ex}4 & $|+1;+1\rangle$ & Bulk & $\sigma^-$ & 23   & 5 & 24 & 3.1 &0\\
  & \rule{0pt}{4ex}11 & $|0; 0\rangle$ & Bulk & $\sigma^+$ & 12 & 0.5 & 9 & 1.8 &-1\\
  & \rule{0pt}{4ex}15 & $|-1;+1\rangle$ & Patches  & $\sigma^+$ & 12 & 5 & 22 & 2.7 &-3 \\
  
  \hline
  
  \multirow{3}{*}{0} & \rule{0pt}{4ex}3 & $|-1;+1\rangle$ & Patches & $\pi$ & 31 &   5  & 21 & 2.7 &4\\
  & \rule{0pt}{4ex}9 & $|0;0\rangle$ & Bulk & $\pi$ & 31  & 0.115 & 14 & 0.8 &0\\
  & \rule{0pt}{4ex}14 & $|-1;+1\rangle$ & Patches & $\pi$ & 31   & 2.5 & 25 & 6&-5  \\
  
  \hline 
  
  \multirow{3}{*}{-1} & \rule{0pt}{4ex}1 & $|-1; +1\rangle$ & Patches & $\sigma^-$ & 8  & 10 & 23 & 3.7&0\\
  & \rule{0pt}{4ex}5 & $|0; 0\rangle$ & Bulk & $\sigma^-$ & 8  & 1 & 11 & 1.5 &1\\
  & \rule{0pt}{4ex}10 & $|-1; -1\rangle$ & Bulk & $\sigma^+$ & 12 &  10 & 21 & 1.8 &-2\\
  
  \hline  
  
  \multirow{2}{*}{-2} & \rule{0pt}{4ex}2 & $| -1; 0\rangle$ & Patches & $\sigma^-$  & 8 &  10 & 13 & 6& -3\\
  & \rule{0pt}{4ex}8 & $|-1; -1\rangle$ & Bulk & $\pi$ & 31& 0.5 & 14 & 2.3 &0\\
  \hline

    -3 & \rule{0pt}{4ex}6 & $|-1; -1\rangle$ & Bulk & $\sigma^-$ & 8 & 5 & 17 & 3.5 &3\\

\end{tabular}
\end{ruledtabular}
\end{center}
\caption{Experimental parameters used to determine the Zeeman diagram of the least-bound rovibrational state. Sub-levels are sorted according to their projection of the total angular momentum $m_F$. We attribute to each level a number ($N$) standing for its position in the energy scale at large magnetic field, $N=1$ being the lowest in energy. Atoms are originally in the $\ket{f = 1, m}$ hyperfine state of the electronic ground state and the different pairs of initial states are labeled with the notation $\ket{m_1; m_2}$. When starting from pure states $\ket{m_1; m_2 = m_1}$, loss spectroscopy is performed on a uniform planar gas while the measurement is done in small patches array for mixtures. The polarization of the microwave field which drives a given atom pair to the dimer state is labeled $P$. We send microwave pulses during a fixed time $t$ for a  given Zeeman level. Their amplitude in the polarization $P$ is given by $B_P$. Two different antennas and microwave sources are used in this work. For each transition we chose the one giving the largest excitation strength. The  orientation of the external magnetic is perpendicular to the plane for all levels but $N=4$, for which we rotate it into the plane to obtain a larger coupling strength. The last three columns give the measured average depletion $\delta n_a/n_a$, the measured FWHM and the distance between the fitted model and the measured lines at a field of 1.25\,G.
\label{tab:zeeman1}}

\end{table*}

\section{Zeeman diagram: experiments}
We show in Table\,\ref{tab:zeeman1} all the relevant experimental parameters used to obtain the Zeeman diagram reported  in Fig.\,3c of the main text, the measured frequency shifts and some information obtained with the model described in the previous section.
\vskip5pt

The statistical uncertainty on the measured positions of the lines is related to the measured linewidth and is typically below 1\,kHz. Systematic uncertainties are dominated by the calibration of the magnetic field $B$. This calibration is realized with well-known atomic transitions and leads to a typical uncertainty on the reported values of the order of 1\,kHz. Note that we have not corrected this measurement with the atom-dimer mean-field interaction. For the M$_1$ transition, this corresponds to a systematic error $\lesssim1\,$kHz and we expect that it is similar for all the other molecular states. In conclusion, all the identified sources of error on the position of the molecular lines in the zero density limit are compatible with an overall accuracy of the order of the kHz.

\end{document}